\documentstyle[psfig]{mn}
\def\H0{{\it H}$_0$}

\def\Ls{{\it L}$_\odot$}
\def\q0{{\it q}$_0$}
\def\kmps{km~s$^{-1}$}

\def\ergps{erg~s$^{-1}$}

\def\nH{$N_{\rm H}$\thinspace} 

\def\psqcm{cm$^{-2}$}
\def\ergpspsqcm{erg~cm$^{-2}$~s$^{-1}$}

\def\Zs{$Z_{\odot}$}

\def\cps{ct\thinspace s$^{-1}$}

\def\phpspsqcm{ph\thinspace s$^{-1}$\thinspace cm$^{-2}$}

\def\pcubcm{cm$^{-3}$}

\title[Hard X-ray emission from Arp220] 
{A hard X-ray constraint on the presence of an AGN in the ultra-luminous infrared galaxy Arp220} 
\author[K. Iwasawa et al] 
{\parbox[]{6.5in} {K. Iwasawa$^1$, G. Matt$^2$, M. Guainazzi$^3$ and A.C. Fabian$^1$}\\
\\
$^1$Institute of Astronomy, Madingley Road, Cambridge CB3 0HA\\ 
$^2$Dipartimento di Fisica, Universita degli Studi Roma Tre, Via della Vasca Navale 84, I-00146 Roma, Italy\\
$^3$XMM-Newton SOC, VILSPA, ESA, Apartado 50727, 28080 Madrid, Spain\\
}
\date{}

\begin{document}

\maketitle

\begin{abstract}
We present X-ray results on the ultraluminous infrared galaxy Arp220
obtained with BeppoSAX.  The X-ray emission up to 10 keV is
detected. No significant signal is detected with the PDS detector in
the higher energy band. The 2--10 keV emission has a flat spectrum
($\Gamma\sim 1.7$), similar to M82, and a luminosity of $\sim 1\times
10^{41}$\ergps. A population of
X-ray binaries may be a major source of this X-ray emission. The upper
limit of an iron K line equivalent width at 6.4 keV is $\simeq 600$
eV. This observation imposes so far the tightest constraint on an
active nucleus if present in Arp220. We find that a column density of
X-ray absorption must exceed $10^{25}$\psqcm\ for an obscured active
nucleus to be significant in the energetics, and the covering factor
of the absorption should be almost unity. The underluminous soft X-ray
starburst emission may need a good explanation, if the bolometric
luminosity is primarily powered by a starburst.
\end{abstract}


\section{introduction}

With an 8--1000$\mu$m luminosity of $L_{\rm ir}\simeq 1.2\times
10^{12}$\Ls, Arp220 is one of the nearest ultra-luminous infrared
galaxies (ULIGs) discovered by IRAS (Soifer et al 1987; we adopt a
distance of 73 Mpc for Arp220 throughout). Since Arp220 is often
regarded as a nearby template for dusty galaxies at high redshift
undergoing a vigorous star formation (e.g., faint SCUBA sources), it
is important to identify its power source. The general consensus since
ISO observations (Sturm et al 1996; Genzel et al 1998) has been massive
young stars. However, some infrared luminous galaxies like NGC4945 and
NGC6240, which apparently show no outward evidence for an active
galactic nucleus (AGN) even in ISO observations, have turned out to
contain a heavily obscured AGN visible only in the hard X-ray band
(e.g., Iwasawa et al 1993; Done et al 1996; Vignati et al 1999). The
soft X-ray data of Arp220 taken by the ROSAT PSPC and HRI show
extended emission similar in shape to the H$\alpha $ nebula.  This can
be well explained by starburst-driven winds (Heckman et al 1996). The
soft X-ray luminosity of $\sim 5\times 10^{40}$\ergps, however,
appears to be small relative to the large infrared luminosity, when
compared with other starburst galaxies. This raises concern with
regard to the intensity of starburst (Iwasawa 1999).

The previous hard X-ray limits on Arp220 by HEAO-1 (Rieke 1988),
CGRO/OSSE (Dermer et al 1997) and Ginga ruled out a bright hard X-ray
source ($>10^{-11}$\ergpspsqcm).  A short ASCA observation also
indicates no evidence for an active nucleus (Iwasawa 1999).  The only
possibility for an energetically significant active nucleus to exist
in Arp220 would be in the form of a ``Compton-thick source'':
no detection of transmitted light in the ASCA bandpass
($\leq 10$ keV) means that a central source must be absorbed by a
column density in excess of $2\times 10^{24}$\psqcm.  Such an X-ray
source could be detected above 10 keV, provided that the column desnity is
not so large, that all the transmitted X-ray flux is highly reduced. 
Although faint
reflected/scattered light may exist, it should be very faint, given
the ASCA constraint on the 5--10 keV emission. A large X-ray
telescope such as Chandra and XMM-Newton is needed to investigate
it.  The main aim of our BeppoSAX observation is to search for a
hard X-ray source above 10 keV with the PDS detector.  Since the ASCA
result indicates that a reflection component alone, if at all present,
should be below the PDS sensitivity limit, any detection with the PDS
would therefore be due to radiation transmitted through a column density
of a few times $10^{24}$\psqcm.


\section{Observation and data reduction}

Arp220 was observed with the NFI instruments (LECS, MECS and PDS) onboard 
BeppoSAX (Boella et al 1997) on two occasions in 2000 August.
Details of the two BeppoSAX observations are given in Table 1.
The LECS and MECS data were reduced from the event files provided by 
the BeppoSAX Data Centre (SDC), using the FTOOLS version 5.0.
The events from the MECS2 and MECS3 have been merged after equalization.
Background data of these imaging instruments were taken from a source-free
region in the observations. These background data were then compared with
the blank sky data extracted from the same detector region 
to correct the source spectra.
The PDS data were reduced via the SAXDAS pipeline processing. 
Significant instrumental background events, which appear as spikes 
in the light curves, have been removed with the 
$5\sigma $ threshould.

We have checked the individual observations and found the two
observations to be consistent in flux and spectral shape. Therefore
the two data sets are merged together for the data analysis below.


\begin{table*}
\begin{center}
\caption{BeppoSAX observations of Arp220. The count rates listed have been 
corrected for background and measured in 
in the 0.1--4 keV for the LECS and 2--10 keV 
for the MECS, and 13--50 keV for the PDS.}
\begin{tabular}{llcc}
Date & Detector & Exposure time & Count rate \\
& & $10^3$ s & $10^{-2}$\cps \\[5pt]
2000 August 4-7 & LECS & 38.3 & $0.19\pm 0.03$ \\
 & MECS & 99.4 & $0.18\pm 0.02$  \\
 & PDS & 47.9 & $0\pm 2$ \\[5pt]
2000 August 27-29 & LECS & 22.5 & $0.17\pm 0.04$ \\
 & MECS & 67.3 & $0.20\pm 0.03$ \\
 & PDS & 29.8 & $5\pm 2$ \\
\end{tabular}
\end{center}
\end{table*}

\section{Results}

\subsection{X-ray images and LECS/MECS spectra}


\begin{figure}
\centerline{\psfig{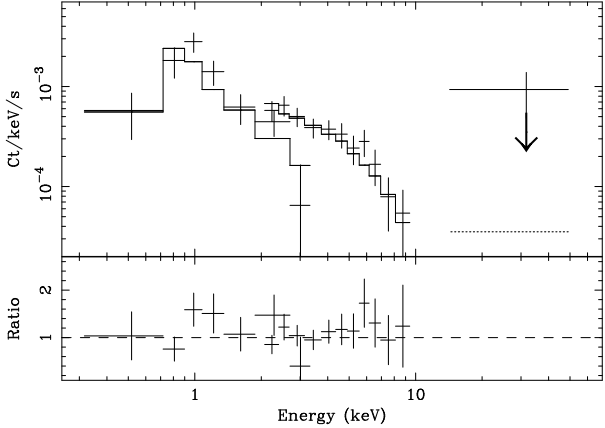}}
\caption{The BeppoSAX spectrum of Arp220 obtained from the LECS, MECS and PDS.
The two observations have been integrated to produce this spectrum. The
histogram shows the best-fit model consisting of an absorbed power-law 
($\Gamma = 1.8$) and 
the MEKAL thermal emision model. The 13--50 keV PDS data point is consistent
with the systematic error of the detector and regarded as an upper limit.
The extrapolation of the best-fit model for the LECS and MECS data is 
indicated by the dotted line. }
\end{figure}


While the short ASCA observation failed to detect X-ray emission above
5 keV, the MECS detects significant X-ray emission up to 10 keV.  The
5--10 keV MECS image shows a point-like source at 7.5$\sigma$ above
the background of the surrounding region. The offset of the position
of this 5--10 keV source from the optical position of Arp220 is 0.7
arcmin, which is within the positional accuracy ($\sim 1$ arcmin) of
BeppoSAX at the time of the observation.

It has been known that a soft X-ray source is located at $\sim 2$
arcmin to the SSW of Arp220 (``southern source'', Heckman et al
1996). The LECS and MECS images below 4 keV show some elongation
towards the southern source. A most likely origin of this source is a
background group of galaxies at a redshift of $\sim 0.09$ (Ohyama et
al 1999). With the relatively broad point spread function of the LECS
and MECS, cross contamination between Arp220 and the southern source
is unavoidable.  We therefore collected spectral data from a region
containing both Arp220 and the southern source (a circular region of a
radius of 3 arcmin for the MECS and 4 arcmin for the LECS,
respectively), and then corrected for the southern source contribution
by utilizing the similarity in spectral shape and flux in the 0.5--2
keV band between the two sources (see Iwasawa 1999; this
spectral similarlity is only valid for a low spectral resolution detector such
as the ROSAT PSPC and a detailed analysis of the ASCA SIS data
does show some difference in temperature and metallicity between the
two sources), which should suffice for the present quality of
data. Above 4 keV, contamination from the southern source is
negligible. The spectrum presented in Fig. 1 contains the southern
source but the flux and luminosity of Arp220 quoted in this paper have
been corrected for the contamination.


A power-law fit to the 2--10 keV MECS data gives $\Gamma =
1.70^{+0.35}_{-0.46}$ when Galactic absorption (\nH = $4\times
10^{20}$\psqcm) is assumed. The spectral slope of Arp220 should be
slightly harder, since the data contains some contribution from the
southern source up to 4 keV. The thermal emisson model requires a temperature
higher than 4.5 keV (the best-fit value is about 10 keV).

Extrapolating the power-law to lower energies leaves surplus soft
X-ray emission below 2 keV, which can be attributed to 
thermal emission from the extended gas. 
Although actual X-ray emission from a starburst galaxy is probably more 
complex than a single temperature plasma (e.g., Dahlem et al 2000),
the present quality of the data is insufficient to decompose
multiple components. The LECS and MECS data were fitted with an
absorbed power-law plus the MEKAL thermal emission model with
Solar abundance, modified only by Galactic absorption.
The temperature of the thermal emission
model obtained is $kT = 0.53^{+0.46}_{-0.22}$ keV. The absorption
column density for the power-law is 
\nH $ = 2^{+4}_{-2}\times 10^{21}$\psqcm, when a photon index of 
$\Gamma = 1.8$ is assumed. The $\chi^2$ value of this fit is 
56.4 for 45 degrees of freedom. 

As a Fe-L bump around 1 keV is evident, metallicity of the gas is not
unusually low, as suggested by the fit to the ROSAT PSPC data by Read
\& Ponman (1998). This is also true for the ASCA SIS spectrum taken
only from the Arp220 region. As far as reasonable metallicity ($Z\geq
0.1$\Zs) is assumed for the MEKAL model, no exess absorption above
Galactic value is required.

No emission line feature is
clearly detected in the spectrum in the energy band above 2 keV; the
excess around 6 keV is not statistically significant.
The 90 per
cent upper limit of a narrow line flux at 6.4 keV in the galaxy-frame
is $1.1\times 10^{-6}$\phpspsqcm, corresponding to the equivalent
width of 600 eV.

The observed fluxes of Arp220 in the 0.5--2 keV and 2--10 keV bands
are $8\times 10^{-14}$ \ergpspsqcm\ and $1.8\times 10^{-13}$
\ergpspsqcm, respectively. These values are consistent with the ASCA
measurements (Iwasawa 1999). The corresponding luminosities for our
assumed distance are $5\times 10^{40}$\ergps\ in the 0.5--2 keV band
and $1.1\times 10^{41}$\ergps\ in the 2--10 keV band. The uncertainty
in the soft X-ray flux induced by the contamination from the southern
source is about 30 per cent at most.

\subsection{PDS upper limit}

The 13--50 keV PDS count rate obtained from the two observations is
$0.032\pm 0.016$ \cps, where the quoted error is statistical only.
This count rate is comparable with the systematic error of the PDS
instrument (Guainazzi \& Matteuzzi 1997), and we therefore treat it as
an upper limit in this energy range. 

\section{interpretation}

\subsection{X-ray emission below 10 keV}


\begin{figure}
\centerline{\psfig{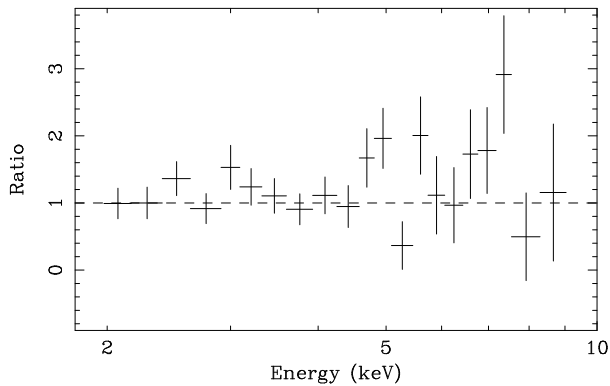}}
\caption{Count rate ratio between Arp220 and M82 obtained from BeppoSAX MECS
observations. The ratios are normalized by the MECS count rate at 2 keV.
The plot is consistent with the hypothesis of a constant ratio over the energy
range ($\chi^2 = 24.2$ for 20 degrees of freedom).}
\end{figure}

The X-ray spectrum of Arp220 below 10 keV can be described by a
combination of sub-keV thermal emission and a moderately absorbed
power-law, as suggested by Iwasawa (1999) based on the ASCA
observation. The soft X-ray emission is explained well with thermal
emission from the interstellar medium (ISM) heated by a Galactic-scale
outflow, probably driven by a starburst (Heckman et al
1996). 

Although the origin of the 2--10 keV emission tail is unclear, we
suggest a population of X-ray binaries as a good candidate.  As shown
in Fig. 2, the 2--10 keV spectra of Arp220 and M82 are very similar.
ASCA monitoring observations of M82 showed a significant X-ray
variability in the 2--10 keV band (Ptak \& Griffiths 1998; Matsumoto
\& Tsuru 1999). A recent Chandra HRC observation revealed that there
are many discrete sources in the galaxy as well as underlying diffuse
hot gas, and the brightest off-nuclear source is highly variable
(e.g., Matsumoto et al 2001). Combining these results, much of the hard
X-ray tail in M82 is likely to be due to a collection of discrete
sources, the majority of which are probably black hole binaries. A similar
picture is also suggested for another nearby starburst galaxy NGC253
with an XMM-Newton observation (Pietsch et al 2001).  A Chandra ACIS
image of the Antennae Galaxy (NGC4038/4039), a galaxy merger with a
strong starburst, also shows many X-ray binaries (Fabbiano, Zezas \& Murray 2001)
which are responsible for the X-ray emission above 2 keV, and thus
support the X-ray binary interpretation.
The X-ray luminosity of the bright
source in M82 in its high state ($8.7\times 10^{40}$\ergps) is
comparable to the luminosity of the 2--10 keV tail in Arp220 ($\sim
1\times 10^{41}$\ergps). 

The lack of a strong Fe K line rules out the possibility of reflected
emission from a hidden active nucleus.  The upper limit of the 6.4 keV
line ($EW\leq 600$ eV) means that any reflection shoud be the order of
half of or less than the observed X-ray emission around 6 keV,
assuming standard cold reflection from optically thick cold matter
with Solar abundance which would have a 6.4 keV line with $EW\simeq 1$
keV (e.g., George \& Fabian 1991). The faint observed hard X-ray
luminosity despite the large bolometric luminosity and suggested high
star formation rate in this galaxy means that the inverse-Compton
scattering, which has been proposed to account for the hard X-ray
emission in starburst galaxies (e.g., Moran, Lehnert \& Helfand 1999),
may not be efficient. However, since this process depends on a number
of parameters we do not know, it cannot be ruled out.

\subsection{Weak soft X-ray emission}


\begin{figure}
\centerline{\psfig{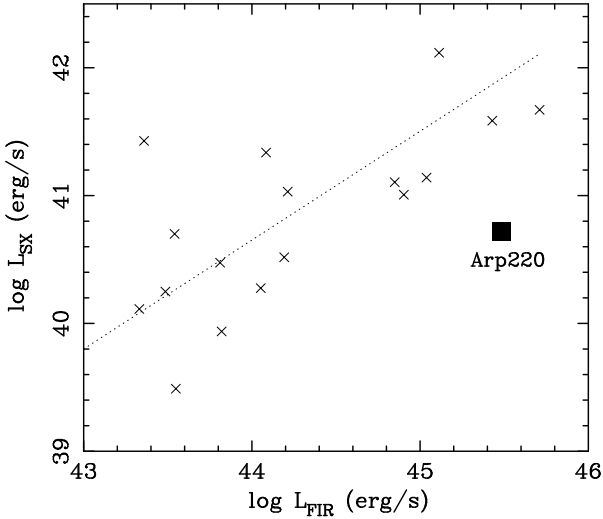}}
\caption{Plot of the 0.1--2 keV luminosity ($L_{\rm SX}$) against
far-infrared luminosity ($L_{\rm FIR}$) in powerful starburst galaxies
(crosses) and Arp220 (filled square). 
The X-ray data obtained either from the ROSAT PSPC or BeppoSAX 
are collected from the literature. The far-infrared luminosities 
are calculated by the formula given in Dahlem et al (1998), based on
the IRAS measurements at 60 and 100$\mu$m from the IRAS Faint Source
Catalogue. The dotted line shows the correlation for the Wolf-Rayet
galaxy sample of Stevens \& Strickland (1998).}
\end{figure}


\begin{figure*}
\centerline{\psfig{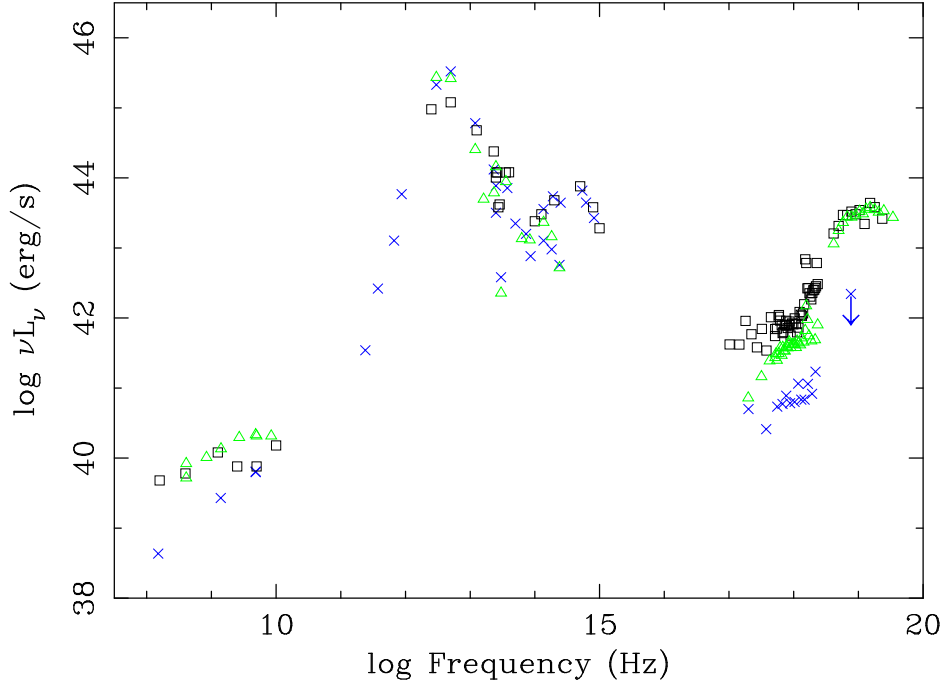}}
\caption{The SED of Arp220 (crosses), NGC6240 (squares), and NGC4945
(triangles). The SED of NGC4945 has been scaled by multiplying by 80.4
for the 60--100$\mu$m luminosity to match Arp220: if NGC4945 were made
to be as powerful as Arp220 at far-infrared wavelengthes, the central source would
have a similar luminosity to that in NGC6240. The soft X-ray decline
of NGC4945 is due to large Galactic absorption (\nH $\sim 1\times
10^{21}$\psqcm). Note that Arp220 is more luminous than NGC6240 in the
far-infrared, but an order of magnitude less luminous in the soft
X-ray band where starburst emission would appear. }
\end{figure*}


The relatively faint soft X-ray emission of Arp220 among powerful
far-infrared galaxies, with some possible reasons, was
discussed in Iwasawa (1999).  Although whether (thermal) soft X-ray 
emission is a good measure of starburst intensity is not entirely clear,
a reasonably good correlation between far-infrared and X-ray luminosities
for star-forming galaxies has been reported (e.g., Griffiths \&
Padovani 1990).
Here we make a further investigation
using other 17 starburst galaxies for which soft X-ray data are 
avaialble. Fig. 3 shows the $L_{\rm
SX}$--$L_{\rm (FIR)}$ relation for far-infrared luminous starburst
galaxies (M82, NGC253, NGC1808, NGC2146, NGC2623, NGC3079, NGC3256,
NGC3310, NGC3628, NGC3690, NGC4038/NGC4039, NGC4631, NGC4666, NGC4945,
NGC6240, Mrk231, and Mrk273) and the correlation obtained for a sample
of nearby Wolf-Rayet galaxies (Stevens \& Strickland 1998). The soft
X-ray (0.1--2 keV) luminosities measured either by the ROSAT PSPC or
BeppoSAX are collected from literature (Dahlem, Weaver \& Heckman
1998; Sansom et al 1996; Zezas, Georgantopolous \& Ward 1998; Heckman
et al 1999; Moran, Lehnert \& Helfand 1999; Awaki et al 1996; Read \&
Ponman 1998; Della Ceca, Griffiths \& Heckman 1997; Della Ceca et al
1999; Turner et al 1997; Vignati et al 1999; Guainazzi et al
2000). The correlation is not very tight, especially in the lower
luminosity range ($L_{\rm FIR}\leq 10^{44}$\ergps) where a
contribution from X-ray binaries becomes significant thus may cause
the scatter in X-ray luminosity. Nevertheless Arp220 appears to be an
order of magnitude underluminous in soft X-rays given the large
far-infrared luminosity.

It is interesting to compare between the wide-band energy
distributions of Arp220 and a similar far-infrared galaxy NGC6240,
which is also a merger system with a comparable, but factor of 2
smaller, infrared luminosity (Fig. 4).  The nuclei of both galaxies
are classified as LINERs (e.g., Veilleux, Kim \& Sanders 1998) and
their optical, near-infrared to mid-infrared spectral features are
dominated by starburst emission (Genzel et al 1998; Lutz, Veilleux \&
Genzel 1999; but also see Dudley 1999). The energy distributions of
the two objects are rather similar from millimetre to optical bands,
but differ significantly in the X-ray band (also in the radio band
longward of several GHz, which could be due to free-free
absorption). Firstly, the soft X-ray emission of Arp220 is about an
order of magnitude below that of NGC6240, as demonstrated in
Fig. 4. Secondly, in NGC6240, a heavily obscured active nucleus
emerges in the hard X-ray band while the soft X-ray emission still
originates from the starburst (Iwasawa \& Comastri 1998; Vignati et al
1999). 
A recent Chandra HRC soft X-ray image of NGC6240 shows no
point-like source but extended emission, suggesting thermal emission
originating from the starburst to be a dominant soft X-ray source,
although it could be a powerful photoionized nebula due to the hidden
active nucleus, reminiscent of that in NGC1068
(Zezas et al 2001).  Both galaxies show large-scale H$\alpha $/soft
X-ray nebulae which are probably both produced by starburst-driven
winds (``superwinds'', Heckman, Armus \& Miley 1990; Heckman et al
1996). The luminosity of the H$\alpha$ nebula of Arp220 is also an
order of magnitude below that of NGC6240 (Heckman, Armus \& Miley
1987). The near-infrared H$_2$ line luminosities of Arp220 and NGC6240
(e.g., Rieke et al 1985) also suggest that, if this line is
shock-heated by the superwinds, the mechanical luminosity of the wind
in Arp220 is a factor of $\sim 10$ smaller than NGC6240. Therefore, in
terms of the products of the superwinds, NGC6240 is much more powerful
than Arp220, and yet possesses a powerful active nucleus hidden from
direct view. 

Besides the X-ray properties, following two points may highlights
possible presence or absense of a massive central objects in the two
galaxies. The stellar velocity dispersions inferred from the
near-infrared CO band absorption features of NGC6240 ($\sigma\sim 360$
km s$^{-1}$) and Arp220 ($\sigma\sim 150$ \kmps) are very different
(Doyon et al 1994; Shier, Rieke \& Rieke 1996). The dispersion
velocity of NGC6240 is, in fact, unusually large compared with other
infrared galaxies. This may indicate that NGC6240 has a massive object
in the centre while Arp220 does not. The other is their radio
morphology.  NGC6240 shows two strong compact sources with diffuse
emission (Colbert et al 1994) whilst many radio sources distributed
over the galaxy, reminiscent of those in the nearby starburst galaxies
M82 and NGC253, have been found in Arp220 (Smith et al 1998).

The nearby far-infrared galaxy NGC4945 is less powerful than 
Apr220 and NGC6240 but has similar characteristics and is a
prototype of active nuclei with absorption column density exceeding
$10^{24}$\psqcm.
Interestingly, the hard X-ray ($\sim $20--150 keV) and
radio emission
of NGC4945 would compare with that of NGC6240, when the SED of NGC4945 
is scaled for the 60--100$\mu $m luminosity 
to match Arp220 (Fig. 4).

Among possible explanations for the small soft X-ray luminosity in
Arp220 are 1) an obscuration in the X-ray nebula; 2) a dense
environment for the star forming region; 3) the age of the starburst;
4) metallicity effects; and 5) a weak starburst. These possibilities
are discussed below.

Observed soft X-ray emission is subject to photoelectric absorption
and could easily be suppresed by an order of magnitude.
Although the soft X-ray emission in Arp220 is extended over a 20 kpc scale
(Heckman, Armus \& Miley 1987), if it is emitted from a number of
small, dusty clumps (see Strickland et al 2000 for the case of
NGC253), suppression of thermal X-ray emission is possible through
absorption within the individual clumps. A study of NaD
absorption features in starburst galaxies by Heckman et al (2000) 
suggests that the column density is a few times of $10^{21}$\psqcm.
However, the absorption feature due to the interstellar medium
is shallower in Arp220 than in NGC6240.
Therefore, as far as the comparison between the two galaxies is 
concerned, there is no reason why the H$\alpha$/X-ray nebula
of Arp220 should be more obscured than NGC6240. Since the extended HI
gas around Arp220 avoids the superwind region (Hibbard, Vacca \& Yun
2000), absorption by this HI gas is also unlikely.
Properties of the obscured starburst region itself are 
constrained by ionizing photon arguments (Shioya, Trentham
\& Taniguchi 2001) and the millimetre emission measurement (Scoville,
Yun \& Bryant 1997). 
We note that the 5--10 keV X-ray emission is least absorbed
and so should be a good direct probe of starburst activity.

To yield the faint superwind features, the mechanical luminosity
emerged outside the starburst region for heating the ISM should be
small (since the superwind-heated soft X-ray luminosity is almost
proportional to the mechanical luminosity, see Heckman et al
1996). Perhaps an unusually dense environment in the starburst region
in Arp220 suppresses the conversion from the mechanical energy of
supernovae and stellar winds to that driving the superwind, due to
less effective thermalization than in other superwind galaxies. The
energy lost here goes to UV radiation via fast cooling of the dense
medium, which, in turn, heats surrounding dust and increases the
infrared emission. This could make a fainter superwind-heated X-ray
nebula and more luminous infrared dust radiation, i.e, a larger
$L_{\rm FIR}/L_{\rm SX}$, than in a lower density medium for which the
initial thermalization is efficient. The condition for this to occur
(cooling time becomes shorter than expansion timescale) would be
the density of the interstellar medium being larger than $\sim 10^6$
cm$^{-3}$ (e.g, Fabian \& Terlevich 1996), which is, however, much
larger than $2\times 10^4$ \pcubcm\ estimated for the nuclear
molecular disk (Scoville et al 1997).

X-ray production by a starburst somewhat delays through thermalization
of the core injection region and formation of X-ray binaries.
The evolution of X-ray emission due to starburst-driven winds is computed
by Strickland \& Stevens (2000) and X-ray properties as a function of
merger evolution are discussed in Read \& Ponman (1998). It is possible
that about one order of magnitude variation in X-ray luminosity occurs after
a burst of star formation. However, since the far-infrared luminosity reflects
on-going star formation, finding a match between large $L_{\rm FIR}$
and a small $L_{\rm SX}$ as seen in Arp220 appears to be difficult.
A study of molecular gas dynamics
suggets that the galaxy merger in NGC6240 is in an earlier stage than
in Arp220 and an even more intensive star formation is still to come in NGC6240
as the merger advances (Tacconi et al 1999). This predicts an increase 
in the far-infrared luminosity of NGC6240 to rival Arp220 but does not
seem to help the soft
X-ray emission in NGC6240, which is already one order of magnitude more
luminous than Arp220, to decline to match that of Arp220, unless 
there is an unknown mechanism to suppress the soft X-ray production 
during the peak of star formation.

Metallicity afftects the emissivity of the hot gas (e.g., Strickland
\& Stevens 2000).
However, the evident Fe-L emission in the soft X-ray spectrum
rules out an anomalously low metallicity in the X-ray nebula.
The massive X-ray binary population appears to depend in part
on the metallicity of a galaxy (Clarke et al 1978). 
This could be due to effects on the initial-mass function, binary evolution
or on stellar wind rates. Unusually high metallicity is needed to suppress
the formation of X-ray binaries.

The last possibility of a weak starburst leaves an accretion-powered
massive black hole to play a significant role in the energetics in
Arp220.  Despite the lack of AGN features in any waveband, this is
still viable, although it is hard to prove.  A central source must be
thoroughly buried in heavy obscuration.  Since the central part of
Arp220 is extremely rich in molecular gas (Sanders, Scoville \& Soifer
1991; Solomon, Radford \& Downes 1992), there is no shortage in supply
for such obscuring material.  Here it is interesting to note that the
dense molecular gas in Arp220 is in the shape of a thin disk with a
radius $\sim 200$ pc containing the two nuclei (Scoville, Yun \&
Bryant 1997), while in NGC6240, the highest molecular concentration
occurs between the two nuclei (Tacconi et al 1999), which may cause a
difference in their nuclear obscuration.  The optical depth of the
dust shrouds must be large so that the temperature of the dust
reradiation we are seeing is cool as observed, and the covering factor
must also be almost unity in order not to expose a hot dust region.
Under this condition, the PAH emission, which is observed strongly 
(Strum et al 1996; Genzel et al 1998), should not be destroyed by X-ray
irradiation (Voit 1992) and the photon-excess problem due to AGN
discussed by Shioya et al (2001) would become irrelevant. A large
covering factor of the obscuration is also required to explain the
lack of reflected X-ray light, i.e., there should be little light
escaping from the nuclear obscuration to be scattered into our line of
sight. One of the few recognizable spectral features expected other
than black body emission is a deep silicate absorption feature at
10$\mu$m, which is indeed observed and a scaled-up protostar model can
explain the mid-infrared spectrum of Arp220 (Dudley \& Wynn-Williams
1997).

Since we detected neither absorbed hard X-ray emission
nor reflection from such an active nucleus, the only viable solution
is to have a Compton-thick source without significant reflected light.
In the next subsection, we use the
PDS upper limit to constrain the properties of such an X-ray nucleus,
if present.

\subsection{Hard X-ray constraint on a hidden active nucleus}

\begin{figure}
\centerline{\psfig{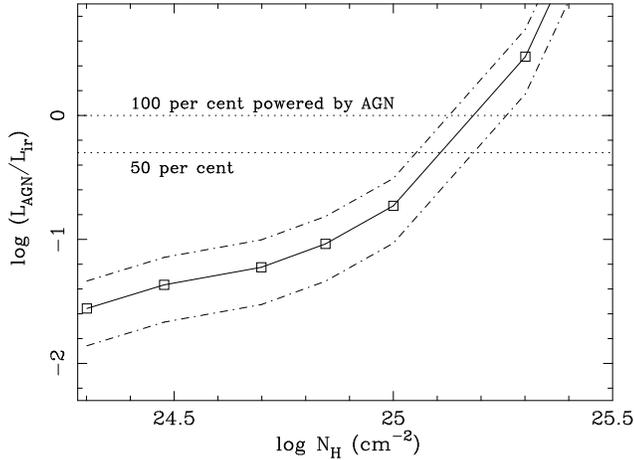}}
\caption{Plot of the upper limit to the 
AGN fraction powering the infrared luminosity of Arp220
as a function of \nH. The solid curve is obtained assuming the ratio of
2--10 keV to bolometric luminosities of the obscured AGN is 5 per cent 
while the dotted-dashed lines are for 3 per cent and 10 per cent.}
\end{figure}

The hard X-ray flux limit implied from the PDS data depends on 
assumed spectral models. In the interest of
an absorbed power-law component, we first estimate the lowest
possible \nH\ value.
Suppose the PDS upper limit was real and no emission from such an absorbed
source is detected with the MECS (see the previous subsection), 
then a lower bound of absorption
column density could be estimated by combining with the MECS data at the
high energy end (say 7--10 keV data).
When a photon index of $\Gamma = 2$ is assumed for the absorbed power-law,
the lower limit of \nH\ is $2\times 10^{24}$\psqcm, where the effect
of Compton scattering in a spherical obscuration is taken into account
(Matt et al 1999). The 13--50 keV flux for the PDS upper limit is then
$3.5\times 10^{-12}$\ergpspsqcm (or $2.1\times 10^{42}$\ergps\ in luminosity).

We then estimate the absorption-corrected 2--10 keV luminosity of the
power-law source with various absorption column densities larger than
the lower limit value obtained above. Spherical obscuration is assumed
for all cases. The 2--10 keV luminosity is then converted to
bolometric luminosity, assuming a typical energy distribution of
Seyfert galaxies and QSOs (e.g., Elvis et al 1994). A reasonable range
of the $L_{\rm 2-10keV}/L_{\rm bol}$ ratio is 3 to 10 per cent.
Provided all the absorbed radiation from an obscured source is
reradiated in the infrared band, the upper limit the AGN fraction
powering the infrared luminosity of Arp220 can be plotted as a
function of absorption column density (Fig. 5). This plot shows that
the column density must exceed $10^{25}$\psqcm\ for a
spherically-obscured active nucleus to be the major power source in
Arp220. In this case, the column density is so high that Compton
down-scattering suppresses the transmitted light to an undetectable
level. As mentioned in the Introduction, even if cold reflection is
present, the limit obtained from the MECS data predicts its flux level
well below the PDS detection limit.

If the hard X-ray source in NGC1068 ($\sim 0.2$ \cps\ detected by
the BeppoSAX PDS, Matt et al 1997), a Compton-thick AGN
with no transmitted hard X-ray component at the distance of 14 Mpc,
is scaled by the $L_{\rm ir}$ and located at the distance of Arp220,
then the expected PDS count rate is $\sim 0.05$\cps, which is
consistent with no significant detection in the PDS data.

\section{summary}

Neither a strong Fe K line nor hard X-ray excess emission is detected 
from Arp220 with BeppoSAX. Although the lack of AGN signatures
points to a starburst being a major energy source of the
luminous infrared emission, the weakness of the soft X-ray emission,
which should also be powered by the starburst, remains puzzling.
In the context of starburst outflow, Arp220 does not appear to
be as powerful as expected from the large infrared luminosity.
An alternative is that much of the bolometric luminosity is powered by
an active nucleus entirely obscured by thick absorbing matter, which
must exceed $10^{25}$\psqcm.

\section*{Acknowledgements}

We thank Dave Sanders, Joss Bland-Hawthorn, Neil Trentham and 
Yoshiaki Taniguchi for useful discussion.
This research has made use of the NASA/IPAC Extragalactic Database
(NED) which is operated by the Jet Propulsion Laboratory, California
Institute of Technology, under contract with the National Aeronautics
and Space Administration. PPARC (KI) and Royal Society (ACF) are 
thanked for support.

\end{document}